\documentclass[a4paper]{article}

\usepackage{INTERSPEECH2018}
\usepackage{multirow}
\usepackage{CJK}
\usepackage{footnote}
\usepackage{array}
\makesavenoteenv{table}
\usepackage[whole]{bxcjkjatype}
\usepackage[unicode]{hyperref}

\title{Multilingual End-to-End Speech Recognition with A Single Transformer on Low-Resource Languages}
\name{Shiyu Zhou$^1$$^,$$^2$, Shuang Xu$^1$, Bo Xu$^1$}
\address{
  $^1$Institute of Automation, Chinese Academy of Sciences \\
  $^2$University of Chinese Academy of Sciences }
\email{\{zhoushiyu2013, shuang.xu, xubo\}@ia.ac.cn}

\begin{document}
\begin{CJK*}{UTF8}{song}

\maketitle

\begin{abstract}
Sequence-to-sequence attention-based models integrate an acoustic, pronunciation and language model into a single neural network, which make them very suitable for multilingual automatic speech recognition (ASR).
In this paper, we are concerned with multilingual speech recognition on low-resource languages by a single Transformer, one of sequence-to-sequence attention-based models.
Sub-words are employed as the multilingual modeling unit without using any pronunciation lexicon.
First, we show that a single multilingual ASR Transformer performs well on low-resource languages despite of some language confusion.
We then look at incorporating language information into the model by inserting the language symbol at the beginning or at the end of the original sub-words sequence under the condition of language information being known during training.
Experiments on CALLHOME datasets demonstrate that the multilingual ASR Transformer with the language symbol at the end performs better and can obtain relatively 10.5\% average word error rate (WER) reduction compared to SHL-MLSTM with residual learning.
We go on to show that, assuming the language information being known during training and testing, about relatively 12.4\% average WER reduction can be observed compared to SHL-MLSTM with residual learning through giving the language symbol as the sentence start token.

\end{abstract}
\noindent\textbf{Index Terms}: ASR, speech recognition, multilingual, low-resource, sequence-to-sequence, Transformer

\section{Introduction}
Multilingual speech recognition has been investigated for many years \cite{huang2013cross,vu2014multilingual,mohan2015multi,sahraeian2016study,zhou2017multilingual}. Conventional studies concentrate on the area of multilingual acoustic modeling by the context-dependent deep neural network hidden Markov models (CD-DNN-HMM) \cite{dahl2012context}. The hidden layers of DNN in CD-DNN-HMM can be thought of complicated feature transformation through multiple layers of nonlinearity, which can be used to extract universal feature transformation from multilingual datasets \cite{huang2013cross}. Among the CD-DNN-HMM based approaches, the architecture of SHL-MDNN \cite{huang2013cross}, in which the hidden layers are shared across multiple languages while the softmax layers are language dependent, is a significant progress in the area of multilingual ASR.
These shared hidden layers and language dependent softmax layers of SHL-MDNN are optimized jointly by multilingual datasets.
SHL-MLSTM \cite{zhou2017multilingual} further explores long short-term memory (LSTM) \cite{hochreiter1997long} with residual learning as the shared hidden layer instead of DNN and achieves better results than SHL-MDNN.

Although these models achieve encouraging results on multilingual ASR tasks, a hand-designed language-specific pronunciation lexicon must be employed. This severely limits their application on low-resource languages, which may have not a well-designed pronunciation lexicon. Recent researches on sequence-to-sequence attention-based models try to remove this dependency on the pronunciation lexicon \cite{sainath2017no,zhou2018comparison,chiu2017state}.
Chiu et al. shows that attention-based encoder-decoder architecture, namely listen, attend, and spell (LAS), achieves a new state-of-the-art WER on a $12500$ hour English voice search task using the word piece models (WPM) \cite{chiu2017state}.
Our previous work \cite{zhou2018comparison} demonstrates that the lexicon independent models can outperform lexicon dependent models on Mandarin Chinese ASR tasks by the ASR Transformer and the character based model establishes a new state-of-the-art character error rate (CER) on HKUST datasets.

Since the acoustic, pronunciation and language model are integrated into a single neural network by sequence-to-sequence attention-based models, it makes them very suitable for multilingual ASR.
In this paper, we concentrate on multilingual ASR on low-resource languages. Building on our work \cite{zhou2018comparison}, we employ sub-words generated by byte pair encoding (BPE) \cite{sennrich2015neural} as the multilingual modeling unit, which do not need any pronunciation lexicon. The ASR Transformer is chosen to be the basic architecture of sequence-to-sequence attention-based model \cite{zhou2018comparison,zhou2018syllable}.
To alleviate the problem of few training data on low-resource languages, a well-trained ASR Transformer from a high-resource language is adopted as the initial model rather than random initialization, whose softmax layer is replaced by the language-specific softmax layer.
We then look at incorporating language information into the model by inserting the language symbol at the beginning or at the end of the original sub-words sequence \cite{li2017multi} under the condition of language information being known during training.
A comparison with SHL-MLSTM \cite{zhou2017multilingual} with residual learning is investigated on CALLHOME datasets with 6 languages.
Experimental results reveal that the multilingual ASR Transformer with the language symbol at the end performs better and can obtain relatively 10.5\% average WER reduction compared to SHL-MLSTM with residual learning. We go on to show that, assuming the language information being known during training and testing, about relatively 12.4\% average WER reduction can be observed compared to SHL-MLSTM with residual learning through giving the language symbol as the sentence start token.

The rest of the paper is organized as follows. After an overview of the related work in Section \ref{label_related work}, Section \ref{label_system_overview} describes the proposed method in detail. We then show experimental results in Section \ref{label_experiment} and conclude this work in Section \ref{label_conclusions}.

\section{Related work}
\label{label_related work}

Although multilingual speech recognition has been studied \cite{huang2013cross,vu2014multilingual,mohan2015multi,sahraeian2016study,zhou2017multilingual} for a long time, these researches are commonly limited to making acoustic model (AM) multilingual, which require language-specific pronunciation model (PM) and language model (LM).
Recently, sequence-to-sequence attention-based models, integrating the AM, PM and LM into a single network, have attracted a lot of attention on multilingual ASR \cite{li2017multi,toshniwal2017multilingual,kim2017towards,watanabe2017language}. \cite{toshniwal2017multilingual,kim2017towards} have presented a single sequence-to-sequence attention-based model can be capable of recognizing any of the languages seen in training. \cite{li2017multi} explored the possibility of training a single model serve different English dialects and compared different methods incorporating dialect-specific information into the model.
However, multilingual ASR on low-resource languages are few investigated by sequence-to-sequence attention-based models.
Furthermore, we argue that the modeling unit of sub-words allows for a much stronger decoder LM compared to graphemes \cite{chiu2017state}, so sub-words encoded by BPE are employed as the multilingual modeling unit rather than graphemes \cite{li2017multi,toshniwal2017multilingual}.

\section{System overview}
\label{label_system_overview}

\subsection{ASR Transformer model architecture}
\label{label_transformer_model}

\begin{figure}[t]
        \centering
        \includegraphics[width=1.0\linewidth]{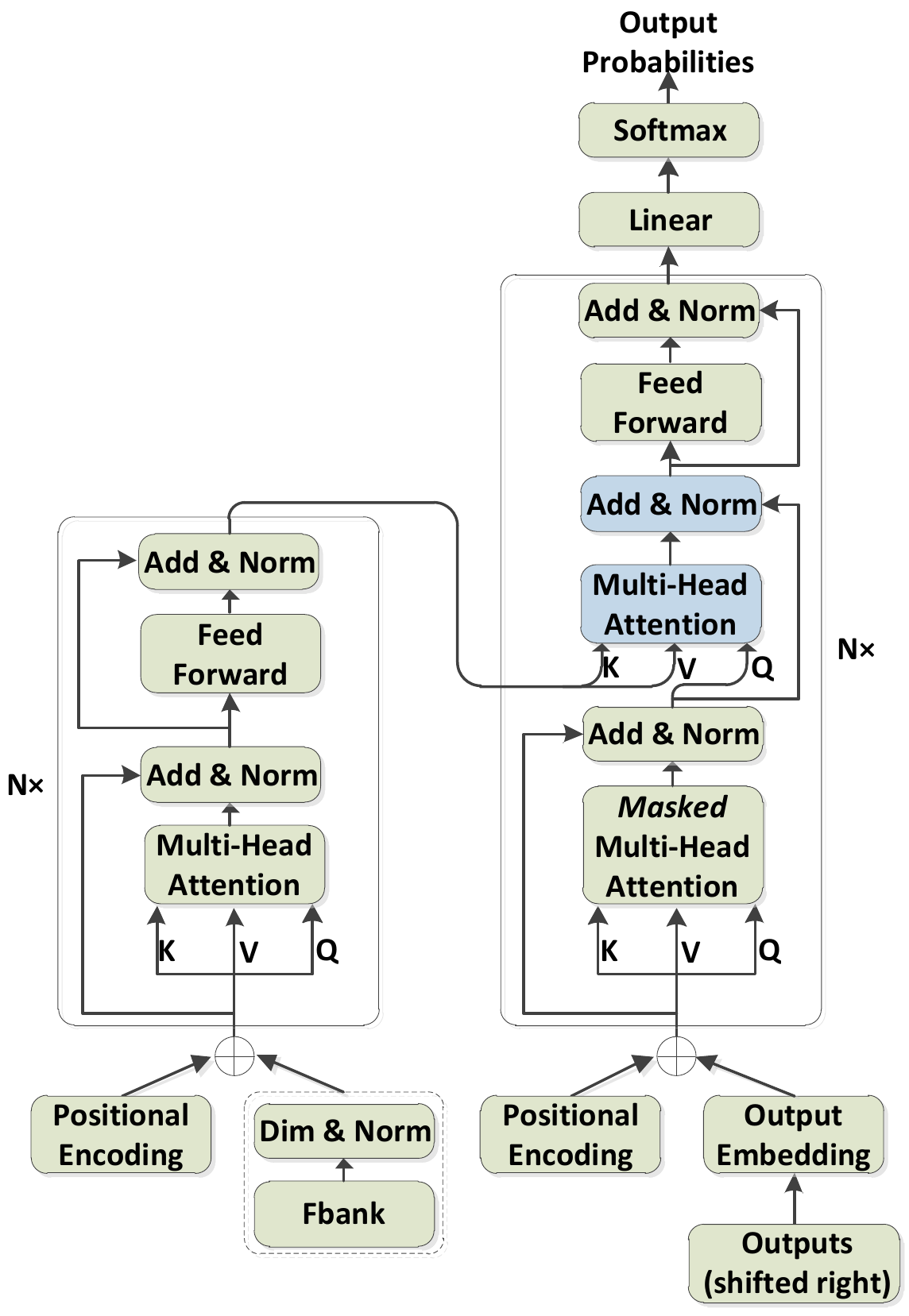}
        \caption{{\it The architecture of the ASR Transformer.}}
        \label{fig:fig_transformer}
        %\vspace{-10pt}
\end{figure}

The ASR Transformer architecture used in this work is the same as our work \cite{zhou2018comparison,zhou2018syllable} which is shown in Figure~\ref{fig:fig_transformer}. It stacks multi-head attention (MHA) \cite{vaswani2017attention} and position-wise, fully connected layers for both the encode and decoder. The encoder is composed of a stack of $N$ identical layers. Each layer has two sub-layers. The first is a MHA, and the second is a position-wise fully connected feed-forward network. Residual connections are employed around each of the two sub-layers, followed by a layer normalization. The decoder is similar to the encoder except inserting a third sub-layer to perform a MHA over the output of the encoder stack. To prevent leftward information flow and preserve the auto-regressive property in the decoder, the self-attention sub-layers in the decoder mask out all values corresponding to illegal connections. In addition, positional encodings \cite{vaswani2017attention} are added to the input at the bottoms of these encoder and decoder stacks, which inject some information about the relative or absolute position of the tokens in the sequence.

The difference between the neural machine translation (NMT) Transformer \cite{vaswani2017attention} and the ASR Transformer is the input of the encoder. We add a linear transformation with a layer normalization to convert the log-Mel filterbank feature to the model dimension $d_{model}$ for dimension matching, which is marked out by a dotted line in Figure~\ref{fig:fig_transformer}.

\subsection{Multilingual modeling unit}
\label{label_multilingual modeling_unit}

Sub-words are employed as the multilingual modeling unit, which are generated by BPE \footnote{https://github.com/rsennrich/subword-nmt} \cite{sennrich2015neural}.
Firstly, the symbol vocabulary with the character vocabulary is initialized, and each word is represented as a sequence of characters plus a special end-of-word symbol `@@', which allows to restore the original tokenization. Then, all symbol pairs are counted iteratively and each occurrence of the most frequent pair (`A', `B') are replaced with a new symbol `AB'. Each merge operation produces a new symbol which represents a character n-gram. Frequent character n-grams (or whole words) are eventually merged into a single symbol. Then the final symbol vocabulary size is equal to the size of the initial vocabulary, plus the \emph{number of merge operations} \emph{$\alpha$} , which is the only hyper-parameter of this algorithm \cite{sennrich2015neural}.

In our multilingual experiments, training transcripts in all languages are combined together to generate the multilingual symbol vocabulary, instead of directly merging each language symbol vocabulary together.
So same sub-words are shared among different languages automatically, which is very beneficial for languages belonging to the same language family. For example, for a German word of ``universitätsgebäu'', it is encoded into ``univer@@ sit@@ ä@@ ts@@ ge@@ b@@ ä@@ u''; for an English word of ``university'', it is encoded into ``univer@@ sit@@ y''. Two sub-words ``univer@@'' and ``sit@@'' are shared in these two languages.

%This can automatically merge the same sub-words from different languages,

\subsection{Language information as output targets}
\label{label_multilingual language_identification}

Similar to \cite{li2017multi,johnson2016google}, we expand the symbol vocabulary of the multilingual ASR Transformer to include a list of special symbols, each corresponding to a language. For example, we add the symbol \textless S\_EN\textgreater\ into the symbol vocabulary when including English.
If the language information of training data can only be known beforehand, two methods of adding the language symbol are explored, i.e. inserting at the beginning (\emph{Transformer-B}) or at the end (\emph{Transformer-E}) of the original sub-words sequence \cite{li2017multi,johnson2016google}.
What's more, if the language information of both training and testing data can be known beforehand, we directly take the language symbol \textless S\_Lang\textgreater\ as the sentence start token (\emph{Transformer-B2}) rather than original sentence start token \textless S\textgreater.
It can force the multilingual ASR Transformer to decode a speech utterance into the pointed language, which is able to alleviate the language confusion greatly during testing.

The difference between \emph{Transformer-B} and \emph{Transformer-B2} is whether to utilize the language information during testing.
The sentence start token is \textless S\textgreater\ in \emph{Transformer-B}. It first predicts a language symbol by itself and then the following tokens are predicted as usual. Therefore, \emph{Transformer-B} do not need to know the language information beforehand during testing.
In contrast, \emph{Transformer-B2} employs \textless S\_Lang\textgreater\ as its sentence start token and predicts the following tokens as usual, which need to know the language information beforehand during testing.
An example of adding the language symbol is shown in Table~\ref{tab:different_language_symbol_examples}.

        \begin{table}[th]
        \caption{\label{tab:different_language_symbol_examples} {\it An example of adding the language symbol.}}
        \vspace{2mm}
        \centerline{
          \begin{tabular}{|c|c|}
            \hline
              {Model}   & {Example}  \\
            \hline
            \hline
            Source &  amazing  \\
            \hline
            Transformer & \textless S\textgreater\ ama@@ z@@ ing\ \textless \textbackslash S\textgreater  \\
            \hline
            Transformer-B & \textless S\textgreater\ \textless S\_EN\textgreater\ ama@@ z@@ ing\ \textless \textbackslash S\textgreater \\
            \hline
            Transformer-E & \textless S\textgreater\ ama@@ z@@ ing\ \textless S\_EN\textgreater\ \textless \textbackslash S\textgreater  \\
            \hline
            Transformer-B2 & \textless S\_EN\textgreater\ ama@@ z@@ ing\ \textless \textbackslash S\textgreater  \\
            \hline
          \end{tabular}
        }
        \end{table}

\section{Experiment}
\label{label_experiment}

\subsection{Data}
The datasets in the paper come from CALLHOME corpora collected by Linguistic Data Consortium (LDC). The following six languages are used: Mandarin (MA), English (EN), Japanese (JA), Arabic (AR), German (GE) and Spanish (SP).
We follow the Kaldi \cite{povey2011kaldi} recipe to process CALLHOME datasets\footnote{the scripts of fisher\_callhome\_spanish in Kaldi are used to process all CALLHOME datasets with some tiny modifications.}.
The detailed information is listed below in Table~\ref{tab:corpora}.
We train the ASR Transformer with a given number of epochs, so validation sets are not employed in this paper.
All experiments are conducted using 80-dimensional log-Mel filterbank features, computed with a 25ms window and shifted every 10ms. The features are normalized via mean subtraction and variance normalization on the speaker basis. Similar to \cite{sak2015fast,kannan2017analysis}, at the current frame $t$, these features are stacked with 3 frames to the left and downsampled to a 30ms frame rate.
We generate more training data by linearly scaling the audio lengths by factors of $0.9$ and $1.1$ \cite{hori2017advances}, since it is always beneficial for training the ASR Transformer \cite{zhou2018comparison}.

      \begin{table}[th]
      \newcommand{\tabincell}[2]{\begin{tabular}{@{}#1@{}}#2\end{tabular}}
        %\vspace{-8pt}
        \caption{\label{tab:corpora} {\it Multilingual dataset statistics.}}
        \vspace{2mm}
        \centerline{
          \begin{tabular}{|c|c|c|c|c|c|c|c|}
            \hline
            {Language}  & \tabincell{c}{\# training utts.}  &  \tabincell{c}{\# test utts.} \\
            \hline
            \hline
               Mandarin (MA)  & $23915$  & $3021$     \\
               English (EN) & $21194$  & $2840$     \\
               Japanese (JA)  & $27165$   & $3381$    \\
               Arabic (AR)  & $20828$  & $2978$    \\
               German (GE)  & $20027$  & $5236$\footnote{We employ devtest as evaltest in German since there is no evaltest from CALLHOME corpora.}  \\
               Spanish (SP)  & $17840$   & $1982$    \\
            \hline
            \hline
               Total  & $130969$   & $19438$    \\
            \hline
          \end{tabular}
        }
      \end{table}

\subsection{Model and training details}

We perform our experiments on the \emph{big model} (D1024-H16) \cite{zhou2018comparison,vaswani2017attention} of the ASR Transformer. Table~\ref{tab:paramters2} lists our experimental parameters. The Adam algorithm \cite{kingma2014adam} with gradient clipping and warmup is used for optimization. During training, label smoothing of value $\epsilon_{ls}=0.1$ is employed \cite{szegedy2016rethinking}. After trained, the last 20 checkpoints are averaged to make the performance more stable \cite{vaswani2017attention}.

At the beginning we train the ASR Transformer on English data with a random initialization, but the result is poor although the CE loss looks good. We propose that one reason for the poor performance could be the training data is too few but the parameters of the ASR Transformer are relatively large which is about $230M$ in this work. To compensate the lack of training data on low-resource languages, a well-trained ASR Transformer with a CER of 26.64\% on HKUST dataset, a corpus of Mandarin Chinese conversational telephone speech, is adopted from our work \cite{zhou2018comparison}.
Its softmax layer is replaced by the language-specific softmax layer which is initialized randomly. Through this initialization method, the ASR Transformer can converge very well. All experiments in this paper are conducted by this initialization method.
        \begin{table}[th]
        \caption{\label{tab:paramters2} {\it Experimental parameters configuration.}}
        \vspace{2mm}
        \centerline{
          \begin{tabular}{|c|c|c|c|c|c|c|}
            \hline
              {model}   & $N$  &    $d_{model}$ & $h$ & $d_k$ & $d_v$   & $warmup$  \\
            \hline
            D1024-H16 & $6$ &   $1024$ & $16$ & $64$ &  $64$  & $12000\ steps$ \\
            \hline
          \end{tabular}
        }
        \end{table}

\subsection{Number of merge operations}
First, we evaluate how the \emph{number of merge operations} \emph{$\alpha$} in BPE affects the performance of the ASR Transformer.
When \emph{$\alpha$} is tiny, the number of sub-words is small. Otherwise the number of sub-words is large.
Since the training data is quite few on low-resource languages, it means that the number of sub-words cannot be too large in order to make sure each sub-word has enough training samples.

For each monolingual ASR Transformer, we first experiment on English dataset for choosing an appropriate \emph{$\alpha$}.
As shown in Table~\ref{tab:different_alpha}, the performance reaches the best when $\emph{$\alpha$}=500$ and the number of sub-words is $548$ on English dataset. Appended with $4$ extra tokens, (i.e. an unknown token (\textless UNK\textgreater), a padding token (\textless PAD\textgreater), and sentence start and end tokens (\textless S\textgreater/\textless \textbackslash S\textgreater)), the total number of sub-words is $552$. In this paper, we choose $\emph{$\alpha$}=500$ in monolingual ASR Transformer experiments.

        \begin{table}[th]
        \caption{\label{tab:different_alpha} {\it WERs(\%) of different $\alpha$ on English dataset.}}
        \vspace{2mm}
        \centerline{
          \begin{tabular}{|c|c|c|c|c|c|}
            \hline
            $\alpha$ & $50$ &   $100$ & $500$ & $1000$ & $2000$ \\
             \hline
            \#\ output. & $106$ &   $156$ & $552$ & $1047$ & $1997$ \\
            \hline
            WER & $45.28$ &   $44.64$ & \textbf{42.77} & $43.88$ & $43.85$ \\
            \hline
          \end{tabular}
        }
        \end{table}

      \begin{table*}[htbp!]
      \newcommand{\tabincell}[2]{\begin{tabular}{@{}#1@{}}#2\end{tabular}}
        %\vspace{-8pt}
        \caption{\label{tab:transformer_results} {\it Comparison of baseline systems and ASR Transformer on CALLHOME datasets in WER/CER (\%). Relative WER/CER reduction is also shown between Multi-Transformer-B2 and SHL-MLSTM-RESIDUAL.}}
        \vspace{2mm}
        \centerline{
          \begin{tabular}{|c|c|c|c|c|c|c|c|c|c|c|}
            \hline
               \multicolumn{2}{|c|}{Model} & {\#\ params.} & {MA} & {EN} & {JA} & {AR} & {GE} & {SP} & {Average} \\
            \hline
            \hline
               \multicolumn{2}{|c|}{Mono-DNN \cite{zhou2017multilingual}} &  $\approx$21.0M & $53.05$ & $50.45$ & $57.52$ & $61.52$ & $59.11$ & $59.77$ & $56.90$   \\
               \multicolumn{2}{|c|}{Mono-LSTM \cite{zhou2017multilingual}} & $\approx$17.8M & $50.53$ & $48.16$ & $55.14$ & $59.21$ & $56.61$ & $57.71$ & $54.56$   \\
               \multicolumn{2}{|c|}{SHL-MDNN \cite{zhou2017multilingual}} & 38.0M & $50.67$ & $46.77$ & $54.15$ & $58.91$ & $55.94$ & $57.88$ & $54.05$   \\
               \multicolumn{2}{|c|}{SHL-MLSTM-RESIDUAL \cite{zhou2017multilingual}} & 22.0M & 45.85 & 43.93 & 50.13 & 56.47 & 51.75 & 53.38 & \textbf{50.25}   \\
            \hline
            \hline
               \multicolumn{2}{|c|}{Mono-Transformer} & $\approx$231M & 39.62 & 42.77 & 39.55 & 50.78 & 48.94 & 54.42 & 46.01   \\
            \hline
              \multirow{4}{*}{Multi} & Transformer & 235M & 40.28 & 42.35 & 39.29 & 50.87 & 47.82 & 53.26 & 45.65   \\
              & Transformer-B & 235M & 40.56 & 41.61 & 38.86 & 50.96 & 47.59 & 53.85 & 45.57   \\
              & Transformer-E & 235M & 40.49 & 40.63 & 38.67 & 50.16 & 47.24 & \textbf{52.58} & \textbf{44.96}   \\
              & Transformer-B2 & 235M & \textbf{37.62} & \textbf{40.36} & \textbf{38.13} & \textbf{48.82} & \textbf{46.22} & 53.07 & \textbf{44.03}   \\
            \hline
              \multicolumn{2}{|c|}{Relative WER/CER Reduction} & $-$ & 17.9\% & 8.1\% & 23.9\% & 13.5\% & 10.7\% & 0.6\% & 12.4\%   \\
            \hline
          \end{tabular}
        }
      \end{table*}

For the multilingual ASR Transformer, all languages training transcripts are combined together to generate the multilingual symbol vocabulary by BPE.
Table~\ref{tab:multilingual_different_alpha} shows that \emph{$\alpha$} do not affect the performance too much on average. We choose $\emph{$\alpha$}=3000$ in all multilingual ASR Transformer experiments and the total number of sub-words is $8062$.

        \begin{table}[th]
        \caption{\label{tab:multilingual_different_alpha} {\it Multilingual results with different $\alpha$ in WER/CER (\%).}}
        \vspace{2mm}
        \centerline{
          \begin{tabular}{|c|c|c|c|c|c|}
            \hline
            $\alpha$ & $1000$ &   $3000$ & $5000$ & $7000$ & $9000$ \\
             \hline
            \#\ output. & $6084$ &   $8062$ & $10025$ & $11959$ & $13883$ \\
            \hline
            MA & 41.14 &   40.28 & 40.66 & 40.14 & 40.72 \\
            EN & 42.76 &   42.35 & 42.49 & 42.73 & 42.76 \\
            JA & 40.04 &   39.29 & 38.63 & 38.68 & 39.76 \\
            AR & 51.04 &   50.87 & 51.32 & 51.15 & 51.80 \\
            GE & 48.92 &   47.82 & 48.85 & 48.21 & 48.11 \\
            SP & 54.34 &   53.26 & 53.07 & 53.37 & 53.73 \\
            \hline
            Average & 46.37 &   \textbf{45.65} & 45.84 & 45.71 & 46.15 \\
            \hline
          \end{tabular}
        }
        \end{table}

\subsection{Results}

        \begin{table}[th]
        \caption{\label{tab:examples_B2} {\it An English example of predictions from Multi-Transformer-B2 with different \textless S\_Lang\textgreater.}}
        \vspace{2mm}
        \centerline{
          \begin{tabular}{|c|c|}
              \hline
              Correct Target & {by any means} \\
              \hline
              \hline
              \textless S\_MA\textgreater\ & {八月 零 零} \\
              \textless S\_EN\textgreater\ & {by any means} \\
              \textless S\_JA\textgreater\ & {バア エリ ミン} \\
              \textless S\_AR\textgreater\ & {tayyib yacni min} \\
              \textless S\_GE\textgreater\ & {war er nicht mit} \\
              \textless S\_SP\textgreater\ & {vaya a dime mil} \\
              \hline
          \end{tabular}
        }
        \end{table}

The baseline systems come from our previous work \cite{zhou2017multilingual} and all results are summarized in Table~\ref{tab:transformer_results}.

First, we train six monolingual ASR Transformers (\emph{Mono-Transformer}) independently on each language data. As can be seen from Table~\ref{tab:transformer_results}, the monolingual ASR Transformer performs very well on each low-resource language and can obtain about relatively 15.7\% WER reduction on average compared to monolingual LSTM (\emph{Mono-LSTM}).

Furthermore, we build a single multilingual ASR Transformer (\emph{Multi-Transformer}) on all training data together without using any language information during training and testing.
We note that the \emph{Multi-Transformer} can achieve slightly better performance than \emph{Mono-Transformer} on average, which represents simply pooling the data together can give an acceptable recognition performance by a single multilingual ASR Transformer.

After analyzing recognition results from \emph{Multi-Transformer}, we find that some recognition results are completely wrong because of language confusion, especially when the speech utterance is short. For example, sometimes an English word ``um'' is decoded into a German word ``ja'', because they have similar pronunciation.

Since the language information of training data usually can be known beforehand, we go on to build two multilingual ASR Transformers integrating language information as depicted in Section \ref{label_multilingual language_identification} to alleviate the problem of language confusion.
Here, the language information is just used during training and the model itself predicts the language symbol during testing. From Table~\ref{tab:transformer_results}, we can observe that inserting the language symbol at the end (\emph{Multi-Transformer-E}) is better than inserting it at the beginning (\emph{Multi-Transformer-B}). Compared to SHL-MLSTM-RESIDUAL, \emph{Multi-Transformer-B} can obtain about relatively 10.5\% average WER reduction.

If the language information of both training and testing data can be known beforehand, we directly take the language symbol \textless S\_Lang\textgreater\ as the sentence start token rather than original sentence start token \textless S\textgreater. It forces the multilingual ASR Transformer to decode a speech utterance into the pointed language, which greatly alleviate the language confusion during testing. As can be seen from Table~\ref{tab:transformer_results}, \emph{Multi-Transformer-B2} performs best and obtain relative 12.4\% average WER reduction compared to SHL-MLSTM-RESIDUAL although the improvement on Spanish is very little. What's more, an interesting observation is that if we give a wrong language symbol \textless S\_Lang\textgreater\ as the sentence start token, \emph{Multi-Transformer-B2} is able to transliterate speech into the pointed language. An English example of predictions from \emph{Multi-Transformer-B2} with different \textless S\_Lang\textgreater\ is shown in Table~\ref{tab:examples_B2}. We can find that the prediction from wrong \textless S\_Lang\textgreater\ is an approximate pronunciation of the correct target.

\section{Conclusions}
\label{label_conclusions}

In this paper we investigated multilingual speech recognition on low-resource languages by a single multilingual ASR Transformer. Sub-words are chosen as the multilingual modeling unit to remove the dependency on the pronunciation lexicon.
A comparison with SHL-MLSTM with residual learning is investigated on CALLHOME datasets with 6 languages.
Experimental results reveal that a single multilingual ASR Transformer by inserting the language symbol at the end can obtain relatively 10.5\% average WER reduction compared to SHL-MLSTM with residual learning if the language information of training data can be employed during training.
We go on to show that about relatively 12.4\% average WER reduction can be observed compared to SHL-MLSTM with residual learning by giving the language symbol as the sentence start token assuming the language information being known during training and testing.
%under the condition of language information being known during training.

\bibliographystyle{IEEEtran}

\bibliography{mybib}

\end{CJK*}
\end{document}